\documentclass[9pt,oneside,reqno]{amsart}
\usepackage{amsfonts,amssymb, amscd,amsmath,latexsym,amsbsy,bm}
\numberwithin{equation}{section}
\usepackage{mathrsfs}
\usepackage{dsfont}
\usepackage{braket}
 \usepackage[usenames,dvipsnames]{xcolor}
 \usepackage[titletoc]{appendix}
\usepackage[normalem]{ulem}

\usepackage[colorlinks,hyperindex, % pagebackref,
           bookmarks=true,bookmarksopen=true]{hyperref}
           \hypersetup    {
        colorlinks=true,
       linkcolor=black,
        urlcolor=MidnightBlue,
        citecolor=RedOrange,
   }

\usepackage{epic}
\usepackage{eepic}
\usepackage{graphicx}

\usepackage{pgf}%,pgfarrows}
\usepackage{tikz}

\usetikzlibrary{shapes}
\usetikzlibrary{plotmarks}

\usepackage{cite}
\usepackage{enumerate}
\usepackage[curve]{xy}
\usepackage{stmaryrd}

\usepackage{mathtools}

\usepackage{mathtext}
\usepackage{stackrel}

\usepackage[utf8]{inputenc}

\begin{document}

$\;$\\
\vspace{2.7cm}

\begin{center}
{\LARGE \bf Emergence of the polydeterminant in QCD}
\end{center}

\vspace{0.2cm}

\vskip 1 cm

\centerline{\large {\bf Francesco Giacosa $^{1,2}$,  Michał Zakrzewski $^{3}$, Shahriyar Jafarzade $^{4,5,6}$ and Robert D. Pisarski $^{7}$}  }

\vskip 0.5 cm
{\small
\begin{center}
\textit{$^1$Institute of Physics, Jan Kochanowski University,
ulica Uniwersytecka 7, P-25-406 Kielce, Poland}\\
\vspace{2.5mm}
\textit{$^2$Institute for Theoretical Physics, Goethe-University,
Max-von-Laue-Straße 1, D-60438 Frankfurt am Main, Germany}\\
\vspace{2.5mm}
\textit{$^3$ Department of Mathematics, Jan Kochanowski University, ulica Uniwersytecka 7, P-25-406 Kielce, Poland}\\
\vspace{2.5mm}
\textit{$^4$ Department of Physics, Arizona State University, Tempe, AZ 85287, USA}\\
\vspace{2.5mm}
\textit{$^5$ Center for Theoretical Physics, Khazar University, Mehseti 41 Street, AZ1096 Baku, Azerbaijan}\\
\vspace{2.5mm}
\textit{$^6$ Composite Materials Research Center, Azerbaijan State Economic University (UNEC), H. Aliyev 135, AZ1063, Baku, Azerbaijan}\\
\vspace{2.5mm}
\textit{$^7$ Department of Physics, Brookhaven National Laboratory, Upton, NY 11973}\\
\end{center}
}

\vskip 0.5cm \centerline{\bf Abstract} \vskip 0.2cm \noindent 

A generalization of the determinant appears in particle physics in
effective Lagrangian interaction terms that model the chiral anomaly in
Quantum Chromodynamics (PRD 97 (2018) 9, 091901 PRD 109 (2024) 7, L071502), in
particular in connection to mesons. This \textit{polydeterminant function}, known in the mathematical literature as a mixed
discriminant, associates $N$ distinct $N\times N$ complex matrices into a
complex number and reduces to the usual determinant when all matrices are
taken as equal. Here, we explore the main properties of the polydeterminant
applied to (quantum) fields by using a formalism and a language close to
high-energy physics approaches. We discuss its use as a tool to write down
novel chiral anomalous Lagrangian terms and present an explicit illustrative model for mesons.
Finally, the extension of the polydeterminant as a function of tensors is shown.

\section{Introduction}
\label{intro}

The determinant is a renowned and essential function in linear algebra that associates an $N \times N$ complex matrix with a complex number, e.g. \cite{strang2009introduction,Zee:2016fuk}. 

Various extensions have been put forward \cite{gelfand1989lectures}, such as the hyperdeterminant (a generalization of the determinant for multidimensional arrays, or tensors) \cite{GELFAND1992226,Ottaviani2013} and the superdeterminant (also known as the Berezinian, which generalizes the determinant for supermatrices in supersymmetric theories), e.g. \cite{Neufeld:1998js,Backhouse:1984vn}. 
Another related concept is the `Pfaffian', which is defined for skew-symmetric matrices and, for even-dimensional matrices, satisfies $(\operatorname{Pf}(A)^2 = \det(A))$ \cite{Forrester2009,WimmerPfaffian}.
 
In this work we concentrate on a different type of generalization of the determinant, denoted in the mathematical literature as `mixed discriminant'.  This was first introduced in 1938 by Alexandrov \cite{Alexandroff1938} by studying mixed volumina, and later on studied by Panov \cite{AAPanov1987} and Bapak \cite{BAPAT1989107} (see also the more recent works \cite{cattani2013mixed,florentin2016characterization,Bapat2015CayleyHamiltonTF}). The mixed discriminant is a function of $N$ distinct $N \times N$ (in general complex) matrices that gives a complex number. When all matrices are set as equal, the usual determinant emerges. Because of this property, it can be regarded as a `polydeterminant' acting on $N$ objects, a term we shall frequently use to describe this function. Notably, this function is connected to the study of $GL(N,\mathbb{C})$ invariants and mixed Cayley Hamilton relations \cite{PROCESI1976306,drensky2005computing,procesi2020tensor}. 
It finds applications within combinatorial studies \cite{ayyer2023combinatorial}, quantum gates \cite{GURVITS2004448}, and information theory \cite{frenkel2015classical}.

Quite interestingly, it turns out that the mixed discriminant /  polydeterminant also appears naturally in the realm of high-energy physics, in particular in Quantum Chromodynamics (QCD). 
In fact, in Ref. \cite{Giacosa:2017pos}  (and in related proceedings \cite{Giacosa:2017ojs}) Giacosa, Pisarski, and Koenigstein (GPK) introduced specific Lagrangian terms when studying certain effective theories of mesons (bound states of a quark and an antiquark that display the symmetries of QCD). In particular, these Lagrangian interaction terms appear when the so-called chiral anomaly (a symmetry of the classical version of QCD broken by quantum fluctuations) is applied to different types of mesons.
Later on, the very same type of Lagrangians have been discussed by Giacosa, Pisarski and Jafarzade (GPJ) in Ref. \cite{Giacosa:2023fdz} by linking its emergence to instantons \cite{Pisarski:2019upw,tHooft:1976snw} , which are non-perturbative Euclidean solutions of the equations of motions of QCD\cite{Belavin:1975fg}. 
%In GPJ a succinct but explicit definition of the mixed discriminant in a qay suitable for setting up Lagrangian terms is given. 

More specifically, the interaction Lagrangian terms described by GKP and GPJ (jointly referred to as GPKJ in the following) are proportional to the polydeterminant mentioned above (yet GPKJ did not notice this point) when the latter is considered as a function of quantum mesonic fields.
It is quite interesting that such a mathematical object enters the description of quantum field theoretical approaches, hence, a deeper study from this point of view seems appropriate. 
In this work, our aim is to discuss the properties and genesis of the polydeterminant in the realm of high-energy physics (HEP) in general and for QCD in particular.

%In the subsequent recent review paper about the extended linear sigma model (a low-energy effective model for QCD) some additional applications of the mixed determinant are listed \cite{Giacosa:2024epf}.

%In fact, the previous papers \cite{Giacosa:2017pos,Giacosa:2023fdz} do not present the details of this function, but merely use it to write down the appropriate interaction Lagrange densities. 
%In fact, the mixed discriminant turns out to be an interesting and versatile tool . Moreover, a rigorous understanding of its features is helpful for future applications in theoretical physics (in QCD and other models of particle physics) as well as further mathematical extensions.

The article is organized as follows: In Sect. 2 the polydeterminant is briefly reviewed by using a formalism typically employed by the HEP community, with special attention to those properties useful for setting up Lagrangian interaction terms, in particular its role as determinant generalization; some useful special cases for $N=2,3$, which are especially important in QCD, are also outlined.  In Appendix A one can find some proofs of the properties listed in Sect. 2. In Sect. 3, we discuss the connection of the polydeterminant to the chiral anomaly in QCD and discuss some applications and examples. Finally, in Sec. 4 we present our conclusions and outlooks. Some more lengthy expressions for the cases $N=4,5$ (also potentially relevant in QCD) are reported in the Appendix B. 

\section{The polydeterminant}

\subsection{Definition and general properties}\hspace{\fill} \\

Given $N$ complex $N\times N$ matrices $A_{1},A_{2},...,A_{N},$ the mixed discriminant or polydeterminant is defined \cite{Alexandroff1938,AAPanov1987,BAPAT1989107} as the function  $\equiv\epsilon:%
%TCIMACRO{\U{2102} }%
%BeginExpansion
\mathbb{C}
%EndExpansion
^{N^{3}}\rightarrow%
%TCIMACRO{\U{2102} }%
%BeginExpansion
\mathbb{C}
%EndExpansion
$ with:%
\begin{equation}
\epsilon(A_{1},A_{2},...,A_{N})=\frac{1}{N!}\epsilon^{i_{1}i_{2}%
...i_{N}}\epsilon^{i_{1}^{\prime}i_{2}^{\prime}...i_{N}^{\prime}}%
A_{1}^{i_{1}i_{1}^{\prime}}A_{2}^{i_{2}i_{2}^{\prime}}...A_{N}^{i_{N}%
i_{N}^{\prime}}%
\text{ ,}
\label{gpkj}
\end{equation}
where the sum is taken over all % $i_{k},i_{k}^{\prime} \in 1,2,\cdots ,N$
 $ i, i' : \{ 1, 2, ..., N \} \to \{ 1, 2, ..., N \} $ and where $\epsilon^{i_{1}i_{2}...i_{N}}$ is
the usual Levi-Civita antisymmetric tensor. 
Another way to express this object
involves two $N$-object permutations $\sigma,\mu$:%
\begin{equation}
\epsilon(A_{1},A_{2},...,A_{N})=\frac{1}{N!}\sum_{\sigma,\mu}\mathrm{sgn} (\sigma
)\mathrm{sgn} (\mu
)A_{1}%
^{\sigma(1)\mu(1)}A_{2}^{\sigma(2)\mu(2)}...A_{N}^{\sigma(N)\mu(N)}\text{ .}%
\label{gpkj2}
\end{equation}
Interestingly, Eq. (\ref{gpkj2}) is much faster when numerical or symbolic calculation is performed.
Note, we prefer here to call the polydeterminant function using $\epsilon(...)$ in order to stress its connection to the Levi-Civita tensors. That is especially important for the chiral anomaly, see Sec. 3.

Below, we list some of the main properties of this object. Their proofs can be found in Appendix A.
%\begin{itemize}

\begin{enumerate}
\item By choosing $A=A_{1}=A_{2}=\cdots =A_{N}$ :
\begin{equation}
\epsilon( A,A,\cdots ,A)=\det\left(  A\right)  \text{ .}%
\end{equation}
Indeed, in the context of GPKJ Lagrangians, this property has been the main motivation behind the construction of the
$\epsilon$-function as a `generalization' of the determinant when $N$ distinct
matrices are involved, see Sec. 3.
In this respect, the interpretation of the $\epsilon$-function as a `polydeterminant' is evident.

\item The $\epsilon$-function is symmetric by exchange of any two matrices:
\begin{equation}
\epsilon ( A_{1},\cdots ,A_{i},\cdots ,A_{j},\cdots A_{N})=\epsilon(
A_{1},\cdots ,A_{j},\cdots ,A_{i},\cdots A_{N})\text{ }%
\end{equation}
for each $i,j=1,\cdots ,N.$

\item The $\epsilon$-function is linear
\begin{align}
\epsilon( A_{1},\cdots , A_{i}= \alpha B_{i} + \beta C_{i},\cdots A_{N})  \, = \
 \qquad \qquad \qquad \qquad  \nonumber \\ \qquad = \alpha \epsilon (
A_{1},\cdots ,B_{i},\cdots A_{N} ) + \beta \epsilon(A_{1},\cdots ,C_{i},\cdots A_{N}) ,
\end{align}
for arbitrary constants $ \alpha , \beta $.

\item By choosing $A=A_{1}$ and $A_{2}=A_3=\cdots =A_{N}=\mathbf{1}$ (where $\mathbf{1}$ is the $N \times N$ identity matrix) the trace emerges:
\begin{equation}\label{mdet-trace}
\epsilon( A,\mathbf{1},\cdots ,\mathbf{1} )=  
\frac{1}{N} \mathrm{Tr}(  A) 
\text{ .}
\end{equation}

\item Upon introducing an invertible matrix $U\in GL(N,%
%TCIMACRO{\U{2102} }%
%BeginExpansion
\mathbb{C}
%EndExpansion
)$ and defining $A_{i}^{\prime}=UA_{i}U^{-1},$ one has:%
\begin{equation}
\epsilon(A_{1}^{\prime},A_{2}^{\prime},\cdots A_{N}^{\prime
})=\epsilon(A_{1},A_{2},\cdots A_{N} )
\end{equation}
A special case is realized for $U$ being an unitary matrix $U(N)$, for which
$A_{i}^{\prime}=UA_{i}U^{\dagger}$. This is the case of `flavor symmetry' in QCD, see Sec. 3.

\item Let $I\subset\{1,2,\cdots ,n\}$ denote any non-empty subset of cardinality
$k$. Then we have
\begin{equation}
\epsilon(  A_{1},A_{2},\cdots ,A_{N})  =\,\frac{1}{N!}\sum
_{I\subset\{1,2,\cdots ,N\}}(-1)^{N-k}\det\left(  \sum_{i\in I}A_{i}\right)  .
\end{equation}

\item  The determinant of the sum of matrices can be written as the sum of each determinant, and $\epsilon (A_{1},A_{2},\cdots ,A_{N})$. More precisely, we can express $ \det(A_1+A_2+\cdots  +A_N) $ as
\begin{align}\label{det-sum}\nonumber
%\det(A_1+A_2+\cdots  +A_N) & = \det(A_1)+\det(A_2)+\cdots  +\det(A_n)+ \\ &
\sum_{\substack{k_1 + k_2 + \cdots + k_r = N \\ k_1, k_2, \ldots, k_r \geq 0}} \binom{N}{k_1, k_2, \ldots, k_r} \epsilon(\underbrace{A_1, A_1, \ldots, A_1}_{k_1}, \underbrace{A_2, A_2, \ldots, A_2}_{k_2}, \ldots, \underbrace{A_r, A_r, \ldots, A_r}_{k_r})  \\
\, = \, \sum_{ k_1 + .. + k_r = N } { N \choose k_1 , k_2 , \cdots  , k_r } \epsilon ( \{ A_1 \}^{k_1} , \{ A_2 \}^{k_2} , \cdots  , \{ A_r \}^{k_r} ) , %\qquad \qquad \qquad \;
\end{align}
where
\begin{equation}\label{}
{ N \choose k_1 , k_2 , \cdots  , k_r } : = \,
 \frac{N!}{k_1! k_2! \cdots  k_r!}
\end{equation}
and where we use the notation\footnote{For example $ ( \{ A \}^3 ) = (A, A, A) $ and $ ( \{ A \}^2 , \{ B \}^3 , \{ C \}^1 ) = (A, A, B, B, B, C) $.}
% $ \{ A_1 \}^{k_1} = A_1, A_1, \cdots , A_1 $, where $ A_1 $ is repeated $ k_1 $ times and so on.
\begin{equation}\label{mulirepl}
 \{ A_1 \}^{k_1} = A_1, A_1, \cdots , A_1
\end{equation}
where $ A_1 $ is repeated $ k_1 $ times, and so on.

\item Upon taking $ N $ matrices $ A_{k} $ and an additional matrix $M$, the following property holds:
\begin{align}
\epsilon(MA_{1},MA_{2},\cdots MA_{N} ) & = \,
\det(M)\epsilon(A_{1},A_{2},\cdots A_{N})
\nonumber
\\ & = \,
\epsilon( A_{1}M,A_{2}M,\cdots A_{N}M)\text{ .}%
\end{align}
This relation can be seen as an extension of the well known identity $ \det \left(A B \right) = \det(A) \det(B)$.

\item 
In general, the $\epsilon$-function can be expressed in terms of traces:%
\begin{equation}
\epsilon(  A_{1},A_{2},\cdots,A_{N})  =\sum_{\substack{n_{1}%
,..,n_{N}\geq0\\n_{1}+2n_{2}+\cdots+Nn_{N}=N\text{ }}}C_{n_{1}n_{2}\cdots
n_{N}}X^{n_{1}n_{2}\cdots n_{N}}%
\end{equation}
with%
\begin{align}
X^{n_{1}n_{2}\cdots n_{N}} &  =\frac{1}{N!}\sum_{\sigma}\mathrm{Tr}\left(
A_{\sigma(1)}\right)  \mathrm{Tr}\left(  A_{\sigma(2)}\right)  \cdot
...\cdot\mathrm{Tr}\left(  A_{\sigma(n_{1})}\right)  \nonumber\\
&  \mathrm{Tr}\left(  A_{\sigma(n_{1}+1)}A_{\sigma(n_{1}+2)}\right)
\mathrm{Tr}\left(  A_{\sigma(n_{1}+3)}A_{\sigma(n_{1}+4)}\right)
\cdot...\cdot\mathrm{Tr}\left(  A_{\sigma(n_{1}+2n_{2}-1)}A_{\sigma
(n_{1}+2n_{2})}\right)  \nonumber\\
&  \mathrm{Tr}\left(  A_{\sigma(n_{1}+2n_{2}+1)}A_{\sigma(n_{1}+2n_{2}%
+2)}A_{\sigma(n_{1}+2n_{2}+3)}\right)  \cdot....\text{ ,}%
\label{Xcoeff}
\end{align}
where the sum refers to all permutations. Above, the term $X^{n_{1}%
n_{2}...n_{N}}$ contains the product of $n_{1}$ traces of a single matrix
$A_{k}$, the product of $n_{2}$ traces of the type $\text{Tr}(A_{k}A_{l})$,
and so on. In particular, it is important to stress that the constraint
\begin{equation}
n_{1}+2n_{2}+\cdots+Nn_{N}=N
\end{equation}
applies. Then, it follows that $n_{N}=0,1$. For $n_{N}=1$ all other entries
vanish.  In general, many terms of the sum of Eq. \ref{Xcoeff} are identical. The coefficients $C_{n_{1}n_{2}\cdots n_{N}}$ follow from the
the Cayley-Hamilton theorem \cite{strang2009introduction}:%
\begin{equation}
C_{n_{1}n_{2}\cdots n_{N}}=\frac{(-1)^{n_{1}+n_{2}+\cdots n_{N}+N}}{1^{n_{1}%
}2^{n_{2}}\cdots N^{n_{N}}n_{1}!n_{2}!\cdots n_{N}!}\text{ .}%
\label{coeff}
\end{equation}
Two simple special cases are given by:%
\begin{align}
X^{N0\cdots0} &  =\mathrm{Tr}\left(  A_{1}\right)  \mathrm{Tr}\left(
A_{2}\right)  \cdots\mathrm{Tr}\left(  A_{N}\right)  \text{ ,}\\
X^{00\cdots1} &  =\frac{1}{N!}\sum_{\sigma}\mathrm{Tr}\left(  A_{\sigma
(1)}A_{\sigma(2)}\cdots A_{\sigma(N)}\right)  \text{ .}%
\end{align}
For $\epsilon(  A,A,\cdots,A) =\det A$ one
recovers the usual expression of the determinant in terms of traces with:
\begin{equation}
X^{n_{1}n_{2}\cdots n_{N}}=\left(  \mathrm{Tr}\left(  A\right)  \right)
^{n_{1}}\left(  \mathrm{Tr}\left(  A^{2}\right)  \right)  ^{n_{2}}%
\cdots\left(  \mathrm{Tr}\left(  A^{N}\right)  \right)  ^{n_{N}}\text{ .}%
\end{equation}
We show the $N=2,3$ specific examples in Sec. 2.2 and the more lengthy expressions for $N=4,5$ in the Appendix B. 
As described in Sec. 3, the introduction of
chiral symmetry and the related chiral anomaly make clear why the $\epsilon$ function is needed in QCD.

\item Geometric meaning: the object $\epsilon(A_{1},\cdots ,A_{i},\cdots ,A_{j}%
,\cdots A_{N})$ is the average of $N!$ oriented volumes of parallelotopes. 
Upon
denoting the matrix $A_{k}$ as%
\begin{equation}
A_{k}=\left(
\begin{array}
[c]{c}%
\mathbf{u}_{k,1}\\
\mathbf{u}_{k,2}\\
\cdots \\
\mathbf{u}_{k,N}%
\end{array}
\right)  \text{ ,}%
\end{equation}
we may rewrite the $\epsilon$-function as:
\begin{equation}\label{geometricspread}
\epsilon( A_{1},\cdots ,A_{N})
=
\sum_{\sigma} \frac{\mathrm{sgn} (\sigma
)}{N!} \mathcal{V(}\mathbf{u}_{\sigma(1),1},\cdots ,\mathbf{u}_{\sigma(N),N}\mathcal{)}
\text{ ,}
%\nonumber
\end{equation}
where $\sigma$ refers to a permutation of $N$ elements, $sgn(\sigma)$ is its
signature, and $\mathcal{V(}\mathbf{u}_{\sigma(1),1},\mathbf{u}_{\sigma
(2),2},\cdots ,\mathbf{u}_{\sigma(N),N})\mathcal{\ }$is the (positive) volume of
the parallelotope spanned by the vectors $\mathbf{u}_{\sigma(1),1}$, $\mathbf{u}_{\sigma(2),2}$, up to $\mathbf{u}_{\sigma(N),N}$.

For example, the
first term is the volume of the parallelotope determined by 
$\mathbf{u}_{1,1},$ $\mathbf{u}_{2,2}$ ... , $\mathbf{u}_{N,N}$. 
For $A_{1}=A_{2}=\cdots =A_{N}=A$, one recovers that $\epsilon ( A,\cdots ,A)=\det(A)$
is the volume of the paralleotope spanned by 
$\mathbf{u}_{1},$ $\mathbf{u}_{2}$,  ... $\mathbf{u}_{N}$,
as expected.
Interestingly, the expression above is the one used to define the mixed discriminant in Ref. \cite{BAPAT1989107}.

%\end{align}

\end{enumerate}
%\end{itemize}

\subsection{The (special) cases $N=2,3$ } \hspace{\fill} \\

Here, we report some of the properties for the simplest non-trivial cases $N=2$ and $N=3$.
These choices appear explicitly in the study of mesonic interactions \cite{Giacosa:2017ojs,Giacosa:2023fdz,Giacosa:2024epf}.

For $N=2$ the explicit expression reads 
\begin{equation}
\epsilon(A,B)=\frac{1}{2}\epsilon^{ij}\epsilon^{i'j'}A^{ii'}B^{jj'}\;.
\end{equation}
In this special case, the previously mentioned properties take the form: 
(1) The determinant emerges as
$\epsilon(A,A)=\det(A)$.
(2) Invariance under exchange: $\epsilon(A,B)=\epsilon(B,A)$.
(3) Linearity: $\epsilon(A_1+A_2,B)=\epsilon(A_1,B)+\epsilon(A_2,B)$.
(4) The trace emerges as
$\epsilon(A,1) = \frac{1}{2}Tr(A)$ ; (5) For $A^\prime=UAU^{-1}$ and $B^\prime=UBU^{-1}$  where $U\in GL(2,\mathbb{C})$ one has $\epsilon(A^\prime,B^\prime)=\epsilon(A,B)$.
(6) Factorization: $\epsilon(MA,MB)=\det(M) \epsilon(A,B)$. 
Points (7) and (8) can be summarized by the following relations:
\begin{equation}    \det(A+B)=\det(A)+\det(B)+2\epsilon(A,B)
\text{ .}
\end{equation}

Point (9) requires a more detailed analysis. The expression in terms of traces reads
\begin{align}
\epsilon(  A_1,A_2 )   &  =C_{20}\text{Tr}\left(  A_1\right)
\text{Tr}\left(  A_{2}\right)  +C_{01}\text{Tr}\left(  A_1A_2\right) \\\nonumber
&  =\frac{1}{2}\left(  \text{Tr}\left(  A_{1}\right)  \text{Tr}\left(  A_{2}\right)
-\text{Tr}\left(  A_{1}A_{2}\right)  \right)
\text{ ,}
\end{align}
with:
\begin{equation}
\text{ }C_{20}=\frac{(-1)^{2+0+2}}{1^{2}2^{0}2!0!}=\frac{1}{2}\text{ ; }%
C_{01}=\frac{(-1)^{0+1+2}}{1^{0}2^{1}0!1!}=-\frac{1}{2}\,.%
\end{equation}

Finally, according to point (10) the geometric interpretation is the average of the area of two parallelograms.

Next, for $N=3$ the explicit form reads: 
\begin{equation}
\epsilon(A,B,C)=
\frac{1}{3!}
\epsilon^{ijk}\epsilon^{i'j'k'}A^{ii'}B^{jj'}C^{kk'} \;.
\end{equation}
The following properties hold: (1) The determinant emerges as
$\epsilon(A,A,A)=\det(A)$. 
(2) Invariance under exchange :$\epsilon(A,B,C)=\epsilon(B,A,C)=\epsilon(C,B,A)$. (3) Linearity: $\epsilon(A_1+A_2,B,C)=\epsilon(A_1,B,C)+\epsilon(A_2,B,C)$.
(4) Trace:
$\epsilon(A,1,1) = \frac{1}{3}\text{Tr}(A)$ ; (5) For $A^\prime=UAU^{-1}$, $B^\prime=UBU^{-1}$, and $C^\prime=UCU^{-1}$
where $U\in GL(2,\mathbb{C})$ implies $\epsilon(A^\prime,B^\prime,C^\prime)=\epsilon(A,B,C)$.
(6) Factorization: $\epsilon(MA,MB,MC)=\det(M) \epsilon(A,B,C)$. 
Points (7) and (8) emerge a special case of Eq. (\ref{det-sum}):
\begin{align}
\det(A+B+C) & = \det(A)+\det(B)+\det(C) + \\ \nonumber
& \qquad 6\epsilon (A,B,C)+3\epsilon (A,A,B) + 3 \epsilon (A,A,C) + \\ \nonumber
& \qquad 3 \epsilon (A,B,B)+3\epsilon (A,C,C)+3\epsilon (B,C,C)+3\epsilon (B,B,C) 
\end{align}
and
\begin{align}
   \epsilon (A, B, C) & = \,
   \det (A + B + C) - \det (A + B) - \det (A + C) - \det (B + C) \nonumber \\
   & \qquad - \det A - \det B - \det C .
\end{align}

\iffalse
It satisfies the following  property
\begin{align}    \epsilon(A,B,C)=\epsilon(B,A,C)=\epsilon(C,B,A)\,,
\end{align}
and reduces to the determinant and trace as follows
\begin{align}    \epsilon(A,A,A)=\det A\,,\qquad \epsilon(A,1,1)=\frac{1}{3}\text{Tr} A\,.
\end{align}
For the non-identity matrices satisfying $A_1\neq A_2\neq A_3$, the following relation works:
\fi
Next, for point (9) we have:
\begin{align}
\epsilon\left(  A,B,C\right)   &  = C_{300}\mathrm{Tr}\left(A\right)  \mathrm{Tr}\left(  B\right)  \mathrm{Tr}\left(  C\right) + \\
\nonumber
& \qquad
\frac{C_{110}}{3}\left(  \mathrm{Tr}\left(  A\right)  \mathrm{Tr}\left(
BC\right)  +\mathrm{Tr}\left(  B\right)  Tr\left(  AC\right)  +\mathrm{Tr}%
\left(  C\right)  \mathrm{Tr}\left(  AB\right)  \right) + \\
\nonumber
& \qquad \frac{C_{001}}{2}\left(  \mathrm{Tr}\left(  ABC\right)  +\mathrm{Tr}%
\left(  ACB\right)  \right)\,,
\end{align}
with
\begin{align}
C_{300}  &  =\frac{(-1)^{3+0+0+3}}{1^{3}2^{0}3^{0}3!0!0!}=\frac{1}{6} \text{ ,} \\
C_{110}  &  =\frac{(-1)^{1+1+0+3}}{1^{1}2^{1}3^{0}1!1!0!}=-\frac{1}{2} \text{ ,} \\
C_{001}  &  =\frac{(-1)^{0+0+1+3}}{1^{0}2^{0}3^{1}0!0!1!}=\frac{1}{3}
\text{ .}
\end{align}
The quantity $\epsilon\left(  A,B,C\right )$ can be rewritten as
\begin{align}
\epsilon\left(  A,B,C\right)   &  =\frac{1}{6}\left[  \mathrm{Tr}\left(
A\right)  \mathrm{Tr}\left(  B\right)  \mathrm{Tr}\left(  C\right)
-\mathrm{Tr}\left(  A\right)  \mathrm{Tr}\left(  BC\right)  -\mathrm{Tr}%
\left(  B\right)  Tr\left(  AC\right)  -\mathrm{Tr}\left(  C\right)
\mathrm{Tr}\left(  AB\right)  +\right. \nonumber \\
&  \left.  \qquad\mathrm{Tr}\left(  ABC\right)  +\mathrm{Tr}\left(
ACB\right)  \right]  \nonumber \text{ ,}
\end{align}
which is rather suggestive and can be easily remembered. See also the Appendix B for more details and for its generalization. 

Finally, point (10) means that the geometric interpretation is the average of the volumes of three parallelepipeds.

There are certain interesting additional relations for the case $N=3$ that we list below. 
\begin{align}
    \epsilon(A,A,B)=\frac{1}{18}\Big(2\det(2A+B)-\det(2B+A)-15 \det(A)+6\det(B)\Big)\,.
\end{align}
From this relation, we derive
\begin{align}
    \epsilon(A,A,1)=\frac{1}{18}\Big(2\det(2A+1)-\det(2\cdot1+A)-15 \det(A)+6\det(1)\Big)
    \text{ .}
    \nonumber
\end{align}
For the traceless matrices A, we obtain the following relation: 
\begin{align}
    \epsilon(A,A,1)=-\frac{1}{3}\text{Tr}(A^2)
\end{align}
\iffalse
In general, it has the following form: 
\begin{align}
   \epsilon(A,A,1)= \frac{1}{3} \Big(-a_{12} a_{21} - a_{13} a_{31} - a_{23} a_{32} + a_{22} a_{33} + a_{11} (a_{22} + a_{33})\Big)
\end{align}
\fi
Moreover:
\begin{align}
   \det (A + B) \, = \,
   \det A + \det B + 3 \left[ \epsilon (A, A, B) + \epsilon (A, B, B) \right] 
   \text{ .}
\end{align}
\iffalse
It follows immediately that %$ \epsilon (A, B, C) $
\begin{align}
   \epsilon (U A U^{-1}, U B U^{-1}, U C U^{-1}) \, = \,
   \epsilon (A, B, C)
\end{align}
and thus that $ \epsilon $ is independent of choice of a coordinate system. 
\fi
The coordinate-system invariance implies also the validity of the formula
\begin{align}
   \epsilon (A, A, 1) \, = \,
   2 (\lambda_1 \lambda_2 + \lambda_1 \lambda_3 + \lambda_2 \lambda_3) \, = \,
   \sigma_2 (\lambda_1 , \lambda_2 , \lambda_3) ,
\end{align}
where $ \lambda_i \in \mathrm{spec} (A) $ and $ \sigma_2 $ denotes \emph{symmetric polynomial} of order\footnote{The other two symmetric polynomials of three variables, $ \sigma_1 (\lambda_1 , \lambda_2 , \lambda_3) : = \lambda_1 + \lambda_2 + \lambda_3 $ and $ \sigma_3 (\lambda_1 , \lambda_2 , \lambda_3) : = \lambda_1 \lambda_2 \lambda_3 $, are associated with the trace and the determinant, respectively.}
$ 2 $.

\section{The polydeterminant in QCD}
In the context of QCD, the mixed discriminant / polydeterminant
function $\epsilon(\cdots )$ arises from the necessity of
incorporating the chiral anomaly into the interactions of mesons. In order to be explicit but without introducing unnecessary details concerning effective mesonic theories and models, let us consider at first, for a given $N_{f}=N$ number of quark flavors\footnote{In QCD one distinguishes the light quark flavors ($u,d,s$) from the heavy ones ($c,b,t$) \cite{ParticleDataGroup:2024cfk}. The case $N_f=N=2$ refers to the quarks $u,d$, while $N_f=3$ to $u,d,s$.}, 
a single $N\times N$ matrix $A$. This matrix contains a multiplet of mesonic fields, such as pions and kaons; see e.g. \cite{Giacosa:2024epf} and refs. therein.
The basic physical requirement is to
construct objects which are invariant by the so-called chiral transformation realizing the famous chiral symmetry, e.g. \cite{Gell-Mann:1960mvl})
\begin{equation}
A\rightarrow U_{L}AU_{R}^{\dagger}\text{ ,}%
\end{equation}
where $U_{L}$ and $U_{R}$ are special unitary matrices belonging to the groups 
$SU(N)_{L}$ and $SU(N)_{R},$ respectively. We recall the matrices $U_L$ and $U_R$ can be expressed as
$U_{L(R)}=e^{-i\theta_{L(R)}^{a}t^{a}}$ with $a=1,\cdots ,N^{2}-1$ with $t^{a}$
being Hermitian traceless matrices that fulfill $\text{Tr}\left(  t^{a}t^{b}\right)
=\frac{1}{2}\delta^{ab}$.
We stress that the matrices $U_L$ and $U_R$ are, in general, distinct from each other. Enforcing them to be equal, $U = U_L = U_R$ we recover the flavor transformation $A \rightarrow U A U^{\dagger}$, see e.g. the textbook \cite{Mosel:1989jf}. Flavor symmetry is an extension of the very famous isospin symmetry postulated long ago by W. Heisenberg \cite{Heisenberg:1932dw} and shortly after formalized by E. Wigner \cite{Wigner:1936dx}.

In the chiral limit (the limit in which the quarks are taken as massless), the classical counterpart of QCD is invariant under the broader symmetry $U(N)_{L}\times
U(N)_{R}$. However, the symmetry under $U(1)_{L}\times U(1)_{R}$ is broken in QCD at the quantum level, resulting in the so-called chiral or axial anomaly \cite{tHooft:1976rip,tHooft:1986ooh,tHooft:1999cta}.
This is a consequence of quantum loops
or, equivalently, of the fact that the interaction measure is, in general, not
invariant under $U(1)_{L}\times U(1)_{R}$ \cite{Fukushima:2017csk}.

Being more specific, a generic $U(1)_{L}\times U(1)_{R}$ transformation amounts to
\begin{equation}
A\rightarrow e^{-i\theta_{L}t^{0}}Ae^{i\theta_{R}t^{0}} = e^{-i\frac{\theta_L - \theta_R}{\sqrt{2N}}}A 
\text{ ,}
\end{equation}
where $\theta_{L,R}$ are the corresponding $U(1)$ group parameters and $t^{0}= \mathbf{1}_{N\times N}/\sqrt{2N}$.
We may express $U(1)_{L}\times U(1)_{R}=U(1)_{V}\times U(1)_{A},$
where $U(1)_{V}$ (in QCD the `baryon number') corresponds to the choice
$\theta_{V}=\theta_{L}=\theta_{R}$ and $U(1)_{A}$ (the so-called chiral transformation) to
$\theta_{A}=\theta_{L}=-\theta_{R}$. 
The transformation under $U(1)_{V}$ reduces to
the identity in the present case. On the other hand, $U(1)_{A}$ leads to the phase transformation
\begin{equation}
A\rightarrow e^{-i\theta_{A}\sqrt{2/N}}A\text{ .}%
\end{equation}
Quantum fluctuations break this transformation. In order
to take this breaking into account, we look for terms that are invariant under $SU(N)_{L}%
\times SU(N)_{R}$ but break $U(1)_{A}.$ It is clear that terms of the type
$\text{Tr}\left(  A^{\dagger}A\right)  ,$ $\text{Tr}\left(  \left(  A^{\dagger}A\right)
^{2}\right)  ,$ $\text{Tr}\left(   (A^{\dagger}A)^3\right)  $ , $\cdots$  are
invariant under $U(N)_{L}\times U(N)_{R}$, thus also under $U(1)_{A}$. These are viable terms that are often used in effective models \cite{Fariborz:2005gm,Parganlija:2012fy,Giacosa:2024epf}.
On the
other hand, the object
\begin{equation}
\det\left(  A\right)
\end{equation}
is invariant under $SU(N)_{L}\times SU(N)_{R}$ but breaks $U(1)_{A}$:%
\begin{equation}
\det\left(  A\right)  \rightarrow\det\left(  e^{-i\theta_{A}\sqrt{2/N}%
}A\right)  =e^{-i\theta_{A}\sqrt{2N}}\det\left(  A\right)  \neq\det\left(
A\right)  \text{ .}%
\end{equation}
This is indeed why the determinant has been used for decades as a viable
description of the chiral anomaly \cite{tHooft:1986ooh}. In particular, when $A$ represents mesonic
fields and is properly embedded into effective Lagrange densities for mesonic
interactions, it leads to a correct phenomenology of the mesons $\eta(545)$
and $\eta^{\prime}(958)$, e.g.  \cite{Rosenzweig:1981cu,Giacosa:2024epf}.
The chiral anomaly via terms involving the determinants is also matter of recent works of the QCD phase diagram  \cite{Pisarski:2024esv,Giacosa:2024orp}. 

An important property concerns the dimension involved in the interaction terms.  We recall that
the dimension of a term in the Lagrangian density amounts to Energy$^4$. 
%It is usually expressed by the product of a parameter $c$ and an operator $O$, $c \cdot O$. The dimension of the operator $O$ corresponds to the number of external legs of the related Feynman diagram.  
Typically, the matrix operator $A$
carries the dimension of Energy$^n$ with $n=1,2,...$. In the simplest case, $n=1$, then $\text{Tr}(A^{\dagger}A)$  scales as Energy$^{2}$ and
$\text{Tr}\left(  \left(  A^{\dagger}A\right)  ^{2}\right)$ as Energy$^{4}$. 
On the
other hand, $\det\left(  A\right)$ scales as Energy$^{N}$. The extension to a different $n$-value is straightforward. 

The question is how to proceed if there is more than a single matrix $A$. This is indeed the case in QCD since different mesonic chiral multiplets do exist, see e.g. the possibilities listed in Ref. \cite{Giacosa:2017pos}.
Let us consider for simplicity two distinct matrices, $A_{1}$ and $A_{2}$, both of
them transforming as $A_{k}\rightarrow U_{L}A_{k}U_{R}^{\dagger}$ with
$k=1,2$. They may refer to distinct mesonic fields, see below for an example. Objects of the type
\begin{equation}
\text{Tr}\left(  A_{1}^{\dagger}A_{2}\right)  \text{ },\text{ }\text{Tr}\left(  \left(
A_{1}^{\dagger}A_{2}\right)  ^{2}\right)  \text{ },\text{ }\text{Tr}\left(
A_{1}^{\dagger}A_{1}A_{2}^{\dagger}A_{2}\right)  \text{ },..
\end{equation}
are chirally invariant under $U(N)_{R}\times U(N)_{L}$ and as such do not break $U(1)_A$.

Our goal is to implement the chiral anomaly when two or more distinct matrices are present. 
One may of course use $\det A_{1}$ as well as $\det A_{2}$ (both with dimension
Energy$^{N}$), as well as their product $\det(A_{1})\det(A_{2}),\cdots $  (with dimension
Energy$^{2N}$). 
%\sout{Yet, that is not the most general way to express chiral anomalous terms. The `anomalous' polydeterminant $\epsilon$-function comes to the rescue.} 
However, this does not represent the most general expression of chiral anomalous terms. The polydeterminant  $\epsilon$-function  provides a tool for systematically constructing chiral anomalous contributions.
In fact, we may consider:
\begin{equation}
\epsilon\left(  A_{1},A_{2},\cdots ,A_{2}\right)
\text{ .}
\end{equation}
If $A_1$ and $A_2$ carry dimension energy, the former term also carries dimension $N$, just as $\det(A_{1})$.
Note, any other combination with one matrix $A_{1}$ and $N-1$ matrices $A_{2}$ is identical to
the one above. Different objects are obtained by considering $N_{1}$ matrices
$A_{1}$ and $N_{2}=N-N_{1}$ matrices $A_{2}$:%
\begin{equation}
\epsilon\underset{N_{1}\text{ times}}{(\underbrace{A_{1},\cdots ,A_{1}}%
}\underset{N_{2}\text{ times}}{,\underbrace{A_{2},\cdots ,A_{2}}})
\text{ .}
\end{equation}
Clearly, $\epsilon\left(
A_{1},,\cdots ,A_{1}\right)  =\det A_{1}$ is obtained for $N_{1}=N$ and $N_{2}=0$ and 
$\epsilon\left(  A_{2},,\cdots ,A_{2}\right)  =\det A_{2}$ for $N_{1}=0$ and $N_{2}=N$.

In the specific case $N=2$, besides $\epsilon\left(  A_{1},A_{1}\right)
=\det\left(  A_{1}\right)  $ and $\epsilon(  A_{2},A_{2})
=\det\left(  A_{2}\right)  $, we have $\epsilon\left(  A_{1}%
,A_{2}\right)$. 
As discussed previously, this quantity can be expressed as
\begin{equation}
\epsilon\left(  A_{1},A_{2}\right)  =\frac{1}{2}\left(  \text{Tr}\left(
A_{1}\right)  \text{Tr}\left(  A_{2}\right)  -\text{Tr}\left(  A_{1}A_{2}\right)  \right)
\text{ .}%
\end{equation}
Each single term of the expression above is \textit{not} invariant neither under
$SU(N)_{R}\times SU(N)_{L}$ nor under $U(1)_{A}$. \textit{Quite remarkably,}
the combination above fulfills $SU(N)_{R}\times SU(N)_{L}$ but breaks
$U(1)_{A}$ just as the determinant does:%
\begin{equation}
\epsilon\left(  A_{1},A_{2}\right)  \rightarrow e^{-i2\theta_{A}%
}\epsilon\left(  A_{1},A_{2}\right)
\text{ .}
\end{equation}

For the case $N=3$ and besides the standard terms $\det(A_{k})$ or their
products such as $\det A_{1}\cdot\det A_{2}$ we might consider
\begin{equation}
\epsilon\left(  A_{1},A_{1},A_{2}\right)  \text{ , }\epsilon\left(
A_{1},A_{2},A_{2}\right)  \text{ .}%
\end{equation}
For isntance, the first one reads:
\begin{align}
&\epsilon\left(  A_{1},  A_{1},A_{2}\right)=  \nonumber 
\\
& \frac{1}{6}
\text{Tr}(A_{1})\left(  \text{Tr}(A_{2})\right)  ^{2}-\frac{1}{6}(\text{Tr}(A_{1})\text{Tr}(A_{2}^2)+2\text{Tr}(A_{2})\text{Tr}(A_{1}A_{2}))
+ \frac{1}{3}\text{Tr}(A_{1}A_{2}^{2})
\text{ .}
\end{align}
Again, each single term is not invariant under $SU(N)_{L}\times SU(N)_{R},$
but the combination above is such. In turn, the chiral anomaly is broken with
\begin{equation}
\epsilon\left(  A_{1},A_{1},A_{2}\right)  \rightarrow e^{-i\theta_{A}%
\sqrt{6}}\epsilon\left(  A_{1},A_{1},A_{2}\right)
\text{ .}
\end{equation}
The case $N=N_{f}=3$ is very useful in practice since there are
three light quark flavors in Nature \cite{ParticleDataGroup:2024cfk}. A Lagrangian term of the type $\epsilon(
A_{1},A_{1},A_{2})  $ contains chirally symmetric but chirally anomalous
interaction terms. 

For the case above, it is useful to present an explicit interaction Lagrangian in
connection to a physically realistic case that serves as an explicit example
of the polydeterminant. To this end, we
note that the complex matrices $A_{1}$ and $A_{2}$ contain 18 real entries
each. A possible connection to physical fields is presented in Ref. \cite{Parganlija:2016yxq}: the
matrix $A_{1}$ describes the ground-state (pseudo)scalar mesons with radial
quantum number $k=1$, while the matrix $A_{2}$ the analogous matrix for the (pseudo)scalar mesons with
radial quantum number $k=2$. In particular, the matrix $A_{k=1,2}$ can be
expressed as
\begin{equation}
A_{k}=%\frac{1}{\sqrt{2}}\sum_{a=0}^{8}\phi_{k}^{a}t^{a}=
\frac{1}{\sqrt{2}}\sum_{a=0}^{8}(s_k^a+ip_k^a)t^a\,,%
\end{equation}
where $s_{k}^{a}=\operatorname{Re}\left[  \phi_{k}^{a}\right]  $ refers to
nine scalar fields and $p_{k}^{a}=\operatorname{Im}\left[  \phi_{k}%
^{a}\right]  $ to  nine pseudoscalar fields for a given radial excitation $k=1,2$.
These fields carry dimension of Energy. Following GPKJ papers, the chirally symmetric but $U(1)_{A}$
anomalous Lagrangian for this system takes the form:
\begin{equation}
\mathcal{L}=c_{1}\det A_{1}+c_{2}\det A_{2}+c_{3}\epsilon(A_{1},A_{1}%
,A_{2})+c_{4}\epsilon(A_{1},A_{2},A_{2})+h.c.
\text{ ,}
\label{lageff}
\end{equation}
where the coefficients $c_{1,2,3,4}$ have also dimension Energy, since
$\mathcal{L}$ must carry Energy$^{4}$ in a four-dimensional world, and $h.c.$
stands for Hermitian conjugate (note, in Ref. \cite{Parganlija:2016yxq} only the first term proportional to $\det(A_1)$
was considered). The first two terms are usual determinants, while the third and the fourth involve the anomalous polydeterminant. The values of the coupling constant is related to instantons \cite{Giacosa:2023fdz}.

We schematically depict the Feynman rules for the four terms of the Lagrangian
of Eq. (\ref{lageff}) in Fig. 1.  
The term proportional to $c_{1}$ is a  standard determinant
\cite{tHooft1976} and implies the self-interaction of the fields of the type
$\phi_{1}^{a}$. 
A related important QCD phenomenological phenomenon linked to this field is the
spontaneous breaking of chiral symmetry, which amounts to rewriting the matrix
$A_{1}$ as:%
\begin{equation}
A_{1}=f_{0}t^{0}+\frac{1}{\sqrt{2}}\sum_{a=0}^{8}\phi_{1}^{a}t^{a}%
\text{ ,}
\label{shift}%
\end{equation}
where $f_{0}$ is a constant: this so-called pion decay constant is
proportional to the quark-antiquark condensate of QCD. It may also be referred to as a vacuum expectation value (v.e.v.). The form above applies
if flavor symmetry is exact (quarks $u$, $d$, and $s$ being exactly massless), what is a
good approximation for our illustrative purposes. The form (or shift) of Eq.
(\ref{shift}) implies that the $c_{1}$-term delivers also quadratic terms of the type $c_{1}f_{0}$ $\left(  p_{1}%
^{a}\right)  ^{2}$, that affect the mass of the mesonic fields. This singlet  mass term $(p_1^0)^2$ plays an important role 
for the already
mentioned properties of the mesons $\eta(547)$ and $\eta^{\prime}(958)$ \cite{ParticleDataGroup:2024cfk}, the
former being closer to $p_{1}^{8}$ and the latter to $p_{1}^{0}$ . 

The term proportional to $c_{2}$ is also a usual determinant and contains analogous three-leg diagrams. If
$A_{2}$ does not undergo condensation, then
this interaction is limited to three-body terms. However, further interaction terms
are possible if $A_{2}$ condenses, see below. 

The third term contains 3-leg terms that mix mesons with $k=1$ and $k=2$. This term makes use of the anomalous polydeterminant. Considering the shift of Eq. (\ref{shift}), other interactions appear, such as the mixing of the type $\phi_{1}%
^{a}\phi_{2}^{b}$ and single-field terms, proportional to $s_{2}^{0}$. In
turn, this term implies that condensation of $A_{2}$ is also possible. This v.e.v. is also anomalously driven (proportional to $c_{3}$)
and mediated by the polydeterminant.

Since it is relevant for showing the importance of the polydeterminant, we explicitly report it here as an example of such novel interaction terms:

\begin{align*}
&\epsilon(A_{1},A_{1}%
,A_{2})
=  \\&4 \sqrt{\tfrac{2}{3}} \, (\phi_{1}^{0})^{2} \phi_{2}^{0}  + \phi_{1}^{4} \left( \tfrac{1}{2} \phi_{1}^{6} \phi_{2}^{1}
- \tfrac{1}{2} \phi_{1}^{7} \phi_{2}^{2}
- \tfrac{1}{2 \sqrt{3}} \phi_{1}^{8} \phi_{2}^{4} \right)  + \phi_{1}^{5} \left( \tfrac{1}{2} \phi_{1}^{7} \phi_{2}^{1}
+ \tfrac{1}{2} \phi_{1}^{6} \phi_{2}^{2}
- \tfrac{1}{2 \sqrt{3}} \phi_{1}^{8} \phi_{2}^{5} \right) \\[4pt]
&- \tfrac{1}{2 \sqrt{3}} \phi_{1}^{6} \phi_{1}^{8} \phi_{2}^{6}
- \tfrac{1}{2 \sqrt{3}} \phi_{1}^{7} \phi_{1}^{8} \phi_{2}^{7}  + \phi_{1}^{2} \left( \tfrac{1}{\sqrt{3}} \phi_{1}^{8} \phi_{2}^{2}
- \tfrac{1}{2} \phi_{1}^{7} \phi_{2}^{4}
+ \tfrac{1}{2} \phi_{1}^{6} \phi_{2}^{5}
+ \tfrac{1}{2} \phi_{1}^{5} \phi_{2}^{6}
- \tfrac{1}{2} \phi_{1}^{4} \phi_{2}^{7} \right) \\[4pt]
&+ \phi_{1}^{1} \left( \tfrac{1}{\sqrt{3}} \phi_{1}^{8} \phi_{2}^{1}
+ \tfrac{1}{2} \phi_{1}^{6} \phi_{2}^{4}
+ \tfrac{1}{2} \phi_{1}^{7} \phi_{2}^{5}
+ \tfrac{1}{2} \phi_{1}^{4} \phi_{2}^{6}
+ \tfrac{1}{2} \phi_{1}^{5} \phi_{2}^{7} \right)  + \phi_{1}^{3} \left( \tfrac{1}{\sqrt{3}} \phi_{1}^{8} \phi_{2}^{3}
+ \tfrac{1}{2} \phi_{1}^{4} \phi_{2}^{4}
+ \tfrac{1}{2} \phi_{1}^{5} \phi_{2}^{5}
- \tfrac{1}{2} \phi_{1}^{6} \phi_{2}^{6}
- \tfrac{1}{2} \phi_{1}^{7} \phi_{2}^{7} \right) \\[4pt]
&+ (\phi_{1}^{8})^{2} \left( -\tfrac{1}{\sqrt{6}} \phi_{2}^{0}
- \tfrac{1}{2 \sqrt{3}} \phi_{2}^{8} \right) + \Big((\phi_{1}^{6})^{2} +(\phi_{1}^{7})^{2} \Big)\left( -\tfrac{1}{\sqrt{6}} \phi_{2}^{0}
- \tfrac{1}{4} \phi_{2}^{3}
- \tfrac{1}{4 \sqrt{3}} \phi_{2}^{8} \right)    \\[4pt]
&+ \Big((\phi_{1}^{4})^{2} +(\phi_{1}^{5})^{2}\Big)\left( -\tfrac{1}{\sqrt{6}} \phi_{2}^{0}
+ \tfrac{1}{4} \phi_{2}^{3}
- \tfrac{1}{4 \sqrt{3}} \phi_{2}^{8} \right)   + \Big( (\phi_{1}^{2})^{2}+\phi_{1}^{3})^{2}\Big) \left( -\tfrac{1}{\sqrt{6}} \phi_{2}^{0}
+ \tfrac{1}{2 \sqrt{3}} \phi_{2}^{8} \right)   \\[4pt]
&- \sqrt{\tfrac{2}{3}} \phi_{1}^{0} \left(
  \phi_{1}^{1} \phi_{2}^{1}
+ \phi_{1}^{2} \phi_{2}^{2}
+ \phi_{1}^{3} \phi_{2}^{3}
+\phi_{1}^{4} \phi_{2}^{4}
+\phi_{1}^{5} \phi_{2}^{5}   
+\phi_{1}^{6} \phi_{2}^{6}
+\phi_{1}^{7} \phi_{2}^{7} +
 \phi_{1}^{8} \phi_{2}^{8}
\right)
\text{ .}
\end{align*}
This expression contains some properties of the determinant, but it mixes $A_1$ and $A_2$. For instance, the first term proportional to $(\phi_{1}^{0})^{2} \phi_{2}^{0} $ is the one responsible for a condensation of $\phi_2^0$.

Finally, the $c_{4}$-term also involves an anomalous polydeterminant and generates mixing and mass terms of the form $\phi_{1}^{a}%
\phi_{2}^{b}$, in particular $c_{4}f_{0}$ $\left(  p_{2}^{a}\right)  ^{2}$. The singlet term with $a=0$ may be relevant to study the resonances $\eta(1295)$ and $\eta(1405)$ \cite{ParticleDataGroup:2024cfk}. 

The detailed phenomenological analysis of these interaction terms goes beyond
the scope of this work (see \cite{Giacosa:2023fdz} for some phenomenological applications), but the arguments above show how the Lagrangian terms that involve the mixed
discriminant can be useful for setting novel and potentially relevant
interaction terms. 

\begin{figure}
  \includegraphics[scale=0.65]{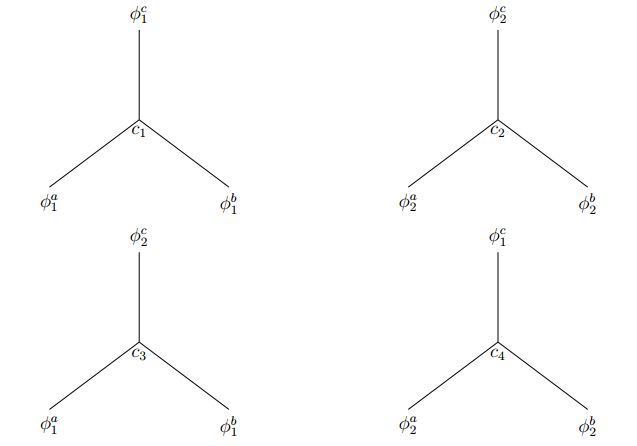}
\caption{Feynman diagrams of Eq. \ref{lageff}. The two upper ones arise from the usual determinant. The two lower ones arise from GPKJ interaction Lagrangians that involve the polydeterminant. Because of that, they involve $\phi_1$ and $\phi_2$ fields: the $c_3$-terms is of the type $\phi_1 \phi_1 \phi_2$ and the $c_4$-terms of the type $\phi_1 \phi_2 \phi_2$.}
\label{fig:1}       % Give a unique label
\end{figure}

 The case $N=4$ may also be of interest in the future (in fact, even if the charm mass strongly breaks chiral symmetry explicitly, as shown in Ref. \cite{Eshraim:2014eka}, certain decay properties still fulfill it.)

Finally, it is important to note that the polydeterminant $\epsilon
(\cdots )$-function can be extended to matrices of fields which carry Lorentz indices. If,
for instance, we have $A_{1\mu}$ as a Lorentz vector and $A_{2\mu\nu}$ as a
Lorentz tensor, the object
\begin{equation}
\epsilon\left(  A_{1\mu},A_{1\nu},A_{2}^{\mu\nu}\right)
\end{equation}
is a Lorentz scalar, provided that the usual Einstein sum is performed. Explicitly:
\begin{equation}
 \sum_{\mu , \nu=0}^{3} \epsilon\left(  A_{1\mu},A_{1\nu},A_{2}^{\mu\nu}\right)
\end{equation}

This property, used in Refs. \cite{Giacosa:2023fdz,Giacosa:2024epf} for practical cases, represents an extension of the polydeterminant to tensors as arguments (instead of plain matrices).
It is evident that upon varying the Lorentz structure and particle types, there
are many Lagrangian terms that can be constructed by using the $\epsilon$ function.

\section{Conclusion}

The Lagrangian terms introduced by GPKJ \cite{Giacosa:2017pos,Giacosa:2024epf} in the study of the chiral anomaly of mesons in QCD make use of a mathematical object known as mixed discriminant that naturally generalized the determinant when different matrices are involved \cite{Alexandroff1938,AAPanov1987,BAPAT1989107}. Hence, we also referred to this object as a polydeterminant. 
Here, we have focused our attention on the main properties of this function that are relevant for the construction of interaction Lagrangian terms in particle physics. 
We have also discussed in detail the connection of this object to effective theories of mesons in QCD and presented an explicit example. Finally, we have shown that the polydeterminant is a suitable determinant generalization for tensors as well.

The GPKJ Lagrangian interaction terms may find other applications besides the ones linked to QCD's effective models, e.g., in the construction of models that go beyond the Standard Model, see e.g. \cite{Lehman:2015via}. Further extensions may consider the inclusion of fermionic objects.

\bigskip

\textbf{Acknowledgments:}
A special acknowledgment goes to A. Koenigstein, with whom Ref. \cite{Giacosa:2017pos} was prepared. 
The authors thank W. Broniowski,T. Brauner,  F. Divotgey, M. Masternak, A. Koenigstein,  P. Kovacs, M. Kyzioł, F. Rennecke, M. Rzeszut,  K. Shekhter, M. Skekalski, and L. Tinti,
for useful discussion. S. J. acknowledges partial support from the U.S. Department of Energy, contract DE-SC0023598.  
R.D.P. was supported by the U.S. Department of Energy under contract DE-SC0012704 and thanks the Alexander v. Humboldt Foundation 
for their support.

\bigskip

\textbf{Data Availability Statement:} This work is purely theoretical, and no datasets were generated or analyzed in this study.

\bigskip

\textbf{Funding and/or Conflicts of Interest/Competing Interests}: The authors declare that there are no conflicts of interest or competing interests associated with this work.

\bigskip

\newpage
\appendix
\section{Proofs of some properties}
In this Appendix, we report, for completeness, the proofs of some of the main properties of the polydeterminant. We follow the same numeration of Sec. 2.

As a first step, we introduce a useful notation for some proofs:
%=
\begin{equation}\label{generalgendet}
\epsilon (A_1, A_2, \cdots , A_N) : =
\frac{1}{N!} \epsilon^{i} \epsilon^{j} A_1^{i_1 j_1} A_2^{i_2 j_2} \cdots  A_N^{i_n j_n} \, ,
\end{equation}
where $ i , j : \{ 1, 2, \cdots  , N \} \to \{ 1, 2, \cdots  , N \} $  and  $\epsilon^i:=\epsilon^{i_{1}i_{2}%
...i_{N}}$.

%\begin{claim}\label{detgendet}
%We have $ \epsilon (A, A, \cdots , A) = \det A $.
%\end{claim}

%\begin{theorem}
%\end{theorem}

\begin{enumerate}
    \item By the very definition of determinant, out of Eq. (\ref{gpkj}) (or, equivalently, from (\ref{gpkj2})) we have $ \epsilon (A, A, \cdots , A) = \det A $.
    \item The function $ \epsilon : \mathbb{C}^{N^3} \simeq M_{N \times N}^N \to \mathbb{C} $ is symmetric and $ N $ linear. Switching $ A_k^{i_k j_k} $ with $ A_l^{i_l j_l} $ in (\ref{generalgendet}) does not modify the value of $ \epsilon $. To adjust to the definition, one has to replace $ i $ and $ j $ with $ i ' $ and $ j ' $, where $ i ' $ and $ j '$ denote the sequences with $ i_k $ interchanged with $ i_l $ and $ j_k $ interchanged with $ j_l $, respectively. But this operation does not change the value of the product $ \epsilon^{i} \epsilon^{j} $, because $ \epsilon^{i} $ and $ \epsilon^{j} $ are either both equal to $ 1 $ or both equal to $ - 1 $, depending on the parity of the permutation. Since transpositions generate the whole permutation group, it follows that the function $ \epsilon $ is symmetric.
    \item Let $ \alpha, \beta \in \mathbb{C} $. Put $ \alpha A + \beta B $ in the first argument, to get
    \begin{align}\label{linearity}
\epsilon (\alpha A + \beta B, A_2, \cdots , A_N) & =
\frac{1}{N!} \epsilon^{i} \epsilon^{j} (\alpha A^{i_1 j_1} + \beta B^{i_1 j_1}) A_2^{i_2 j_2} \cdots  A_N^{i_N j_N} \nonumber \\
 & = \, \alpha \frac{1}{N!} \epsilon^{i} \epsilon^{j} A^{i_1 j_1} A_2^{i_2 j_2} \cdots  A_N^{i_N j_N} \nonumber \\ & \qquad + \beta \frac{1}{N!} \epsilon^{i} \epsilon^{j} B^{i_1 j_1} A_2^{i_2 j_2} \cdots  A_N^{i_N j_N} \nonumber \\
 & = \, \alpha \epsilon (A, A_2, \cdots , A_N)  + \beta \epsilon (B, A_2, \cdots , A_N) .
\end{align}
Thus $ \epsilon $ is linear w.r.t. the first argument. But we know that $ \epsilon $ is symmetric, so it's linear w.r.t. any argument and hence $ N $-linear. The %theorem
 property is proved. \; $ \square $

\item The formula (\ref{mdet-trace}) follows from straightforward computation.

%\begin{theorem}
\item Function $ \epsilon : \mathbb{C}^{N^3} \to \mathbb{C} $ is invariant with respect to the choice of basis.
%\end{theorem}

Algebraically, the invariance with respect to the choice of basis is represented by the fact that the value of $ \epsilon (A_1, A_2, \cdots , A_N) $ is the same as the value of $ \epsilon (U A_1 U^{-1}, U A_2 U^{-1}, \cdots , U A_N U^{-1}) $, where $ U \in GL (N, \mathbb{C}) $. In coordinates, we have
\begin{equation}
    (U A_k U^{-1})^{i_l j_l} \, = \,
    U_{i_l i_l '} A_k^{i_l ' j_l '} U^{j_l ' j_l} ,
\end{equation}
whereby $ U^{j_l ' j_l} $ we denote the entrances of the inverse matrix $ U^{-1} $ and by $ U_{i_l i_l '} $ we denote the entrances of the matrix $ U $. Let us denote
\begin{equation}
\epsilon ' \, = \,
    \epsilon (U A_1 U^{-1}, U A_2 U^{-1}, \cdots , U A_N U^{-1}) .
\end{equation}
Now, plugging in the conjugated matrices into (\ref{generalgendet}), we get
\begin{align}\label{generalgendetconj}
     \epsilon ' & : =
\frac{1}{N!} \epsilon^{i} \epsilon^{j} U_{i_1 i_1 '} A_1^{i_1 j_1} U^{j_1 ' j_1}  
 \cdots  U_{i_N i_N '} A_N^{i_N j_N} U^{j_N ' j_N} \nonumber \\ \,
 & : =
\frac{1}{N!} \epsilon^{i} \epsilon^{j} U_{i_1 i_1 '} \cdots  U_{i_N i_N '} A_1^{i_1 j_1} %A_2_{i_2 j_2}
 \cdots  A_N^{i_N j_N} U^{j_1 ' j_1} \cdots  U^{j_N ' j_N} \nonumber \\
 & : =
\frac{1}{N!} \det U \epsilon^{i '} \epsilon^{j} U_{i_1 i_1 '} \cdots  U_{i_N i_N '} A_1^{i_1 j_1} %A_2_{i_2 j_2}
 \cdots  A_N^{i_N j_N} \det U^{-1} \epsilon^{i '} \nonumber \\
 & : = \, \epsilon (A_1, A_2, \cdots , A_N) \, ,
\end{align}
where in the third line we used the formulas\footnote{See e.g. page 941 of "Mathematical methods for physics and engineering" by K. F. Riley, M. P. Hobson and S. J. Bence.}
\begin{equation}\label{}
\epsilon^{i} U_{i_1 i_1 '} \cdots  U_{i_N i_N '} \, = \,
 \det U \epsilon^{i '}
\end{equation}
and
\begin{equation}\label{}
\epsilon^{j} U^{j_1 ' j_1} \cdots  U^{j_N ' j_N} \, = \,
 \det U^{-1} \epsilon^{j} .
\end{equation}
The %theorem
 result is proved. \; $ \square $
%\\

\item From the properties (\ref{linearity}) and (\ref{detofsum}), it follows that the function $ \epsilon $ is essentially associated to commutative algebra (despite matrices, or  operators, being generally non-commutative) and as such, it resembles many properties of symmetric tensors. This manifests, for example, in the possibility of expressing $ \epsilon $ in terms of combination of determinants. In particular, we have the following formula.

Let $ I \subset \{ 1, 2, \cdots , N \} $ denote any nonempty subset of cardinality $ k $. Then we have
\begin{equation}\label{epsilasdets}
\epsilon (A_1, A_2, \cdots , A_N) \, = \, 
 \frac{1}{N!} \sum_{I \subset \{ 1, 2, \cdots , N \} } (-1)^{N - k} \det \left( \sum_{i \in I} A_i \right) .
\end{equation}

The proof of identity (\ref{epsilasdets}) is based on the following observation. Let $ x_1, x_2, \cdots , x_N $ denote commuting variables. Then the expression
\begin{equation}\label{epsilasdets'}
\sum_{I \subset \{ 1, 2, \cdots , N \} } (-1)^{N - k} \left( \sum_{i \in I} x_i \right)^N ,
\end{equation}
is divisible by $ x_i $ for all $ i = 1, 2, \cdots , N $. To see this, let us first rewrite (\ref{epsilasdets'}) as
\begin{equation}\label{}
%\sum_{I \subset \{ 1, 2, \cdots , n \} } (-1)^{n - k} \left( \sum_{i \in I} x_i \right)^n \, = \,
 \sum_{I \subset \{ 2, \cdots , N \} } (-1)^{N - k - 1} \left( x_1 + \sum_{i \in I} x_i \right)^N + \sum_{I \subset \{ 2, \cdots , N \} %\not \ni 1
 } (-1)^{N - k} \left( \sum_{i \in I} x_i \right)^N ,
\end{equation}
where now $ I \subset \{ 2, \cdots , N \} $ is a subset of $ k $ elements, where $ k = 1, 2, \cdots , N - 1 $. Indeed, this equation is simply the identity $N!\,x_1\cdots x_N$, written in symmetric form.  
For example, for $N=2$,
$2!\,x_1x_2=(x_1+x_2)^2-x_1^2-x_2^2$,  
and for $N=3$,
$3!\,x_1x_2x_3=(x_1+x_2+x_3)^3-(x_1+x_2)^3-(x_1+x_3)^3-(x_2+x_3)^3+x_1^3+x_2^3+x_3^3$.
Now, if we put $ x_1 = 0 $, we get
\begin{equation}\label{}
 \sum_{I \subset \{ 2, \cdots , N \} } (-1)^{N - k - 1} \left(\sum_{i \in I} x_i \right)^N + \sum_{I \subset \{ 2, \cdots , N \} } (-1)^{N - k} \left( \sum_{i \in I} x_i \right)^N \, = \, 0 ,
\end{equation}
so (\ref{epsilasdets'}) is clearly divisible by $ x_1 $. But the expression (\ref{epsilasdets'}) is symmetric w.r.t. all variables $ x_1, x_2, \cdots , x_N $, so we get the result. \; $ \square $
\\

\textit{Proof of the property}. Using the principle of commutativity and linearity, we can plug in matrix $ A_i $ in place of variable $ x_i $, to get the result.  \; $ \square $
%\\

\item
%\begin{theorem}
Function $ \epsilon : \mathbb{C}^{N^3} \simeq M_{N \times N}^N \to \mathbb{C} $ satisfies the combinatorial formula
\begin{align}\label{detofsum}
\det (A_1 + A_2 + \cdots  + A_N) %& =
\qquad \qquad \qquad \qquad \qquad \qquad \qquad \qquad
 \nonumber \\
 %\qquad \qquad \qquad \qquad \qquad \qquad
 \, = \, \sum_{ k_1 + .. + k_r = N } { N \choose k_1 , k_2 , \cdots  , k_r } \epsilon ( \{ A_1 \}^{k_1} , \{ A_2 \}^{k_2} , \cdots  , \{ A_r \}^{k_r} ) ,
\end{align}
where
\begin{equation}\label{}
{ N \choose k_1 , k_2 , \cdots  , k_r } : = \,
 \frac{N!}{k_1! k_2! \cdots  k_r!}
\end{equation}
and where we again use the notation as in (\ref{mulirepl}).  
Using the  
 first property and putting $ A_1 + A_2 + \cdots  + A_N $ as an argument of determinant, we see that it is invariant w.r.t. permutations. Thus the combinatorics of $ \det (A_1 + A_2 + \cdots  + A_N) $ expanded with use of (\ref{generalgendet}), obeys the same law as usual power $ (A_1 + A_2 + \cdots  + A_N)^N $. The %theorem
 property is proved. \; $ \square $

\item Another property (which again resembles similarity to commutative algebra, instead of modules over a non-commutative ring) is the following. If $ M \in \mathbb{C}^{N^2} $ is any matrix, then
\begin{align}\label{}
\epsilon (M A_1, M A_2, \cdots , M A_N) & = \, 
 \det M \cdot \epsilon (A_1, A_2, \cdots , A_N) \nonumber \\ & = \,
 \epsilon (A_1 M, A_2 M, \cdots , A_N M) .
\end{align}

%\begin{rem}
    The proof can be given directly, but we can also use the commutativity and linearity of $ \epsilon $ to get the result in much simpler way.
%\end{rem}

\textit{Proof}: If $ \epsilon (A_1, A_2, \cdots , A_N) $ corresponds to the product of the formal variables $ x_1 x_2 \cdots  x_N $ and $ M $ corresponds to a formal variable $ y $, then 
\[ y x_1 y x_2 \cdots  y x_N = y^N x_1 x_2 \cdots  x_N . \]
But $ y^N $ corresponds to the determinant $ \det M $. Since the multiplication of the formal variables $ x_1 , x_2 , \dots , x_N $ and $ y $ is commutative, the result follows.  \; $ \square $

\item 
The expression of $\epsilon(A_1,...,A_N)$ in terms of traces is a direct consequence of property (7), upon rewriting the determinants in terms of traces. The coefficients of Eq. (\ref{coeff}) coincide with the Cayley-Hamilton theorem ensuring that the usual expression for $ \det (A_1) $ emerges when $A_1 =...=A_N$ is set. The property follows from the requirement of point (2), i.e. invariance under the exchange of arbitrary entries. This property has also been numerically verified for $N=2,3,4,5$ (for the latter two cases see Appendix). 

\item One of the classical definitions of determinant (see e.g. \cite{gelfand1989lectures}) is given by
\begin{equation}\label{clasicdet}
\det A \, = \,
\sum_{\sigma} \mathrm{sgn} (\sigma) A^{1 ,\sigma(1)} A^{2 ,\sigma(2)} \cdots A^{N ,\sigma(N)} , 
\end{equation}
with $ \sigma $ running over all permutations of the set $ \{ 1, 2, ..., N \} $. Recalling that the oriented volume of the set of $ N $ vectors in $ N $-dimensional space is given by determinant and plugging in (\ref{clasicdet}) into (\ref{geometricspread}), we get
\begin{align}
\sum_{\sigma} \frac{\mathrm{sgn} (\sigma)}{N!} \mathcal{V(}u_{\sigma(1),1,},\cdots ,u_{\sigma(N),N}\mathcal{)} & = \,
\nonumber \\
%\qquad
\frac{1}{N!} \sum_{\sigma} \mathrm{sgn} (\sigma) \left( \sum_{\tau} \mathrm{sgn} (\tau) A_{\sigma (1) }^{1 , \tau (1)} A_{\sigma (2) }^{2 , \tau (2)} \cdots A_{\sigma (n) }^{n , \tau (n)} \right) & = \, \nonumber \\
%\qquad
\frac{1}{N!} \sum_{\sigma , \tau} \mathrm{sign} ( \sigma ) \cdot \mathrm{sign} ( \tau ) %\qquad \qquad \qquad \qquad & \nonumber \\ \cdot
A_{1 }^{\sigma^{-1} (1) , \tau (1)} A_{2}^{\sigma^{-1} (2) , \tau (2)} \cdots A_{n}^{\sigma^{-1} (n) , \tau (n)} & = \, \nonumber \\
 \frac{1}{N!} \sum_{i , j} \epsilon^i \cdot \epsilon^j A_{1}^{i_1 , j_1} A_{2}^{i_2 , j_2} \cdots A_{n}^{i_n , j_n} & = \,
 \epsilon (A_1, ..., A_N) , \nonumber
%\text{ ,}
%\nonumber
\end{align}
where, in the last line, we used the fact that $ \epsilon^i $ corresponds to the sign of the permutation given by $ (1, 2, ..., N) \to (i (1), i (2), ..., i (N)) $.  \; $ \square $

\end{enumerate}

\iffalse
$$ \cdots  $$

\begin{itemize}
\item {\color{BrickRed}What is}: $ \epsilon (A, A, I , \cdots , I) $, $ \epsilon (A, A, A, I , \cdots , I) $, and so on$\cdots$  (symmetric polynomials of eigenvalues?). {\color{BrickRed} It is true, but there is no general proof yet}:
\item Is it true, in general that
\begin{equation}\label{}
\epsilon (U A_1 U^{-1}, U A_2 U^{-1}, \cdots , U A_N U^{-1}) =
\epsilon (A_1, A_2, \cdots , A_N) , 
\end{equation}
for any $ U \in GL (n , \mathbb{C}) $? This would be a very strong property - invariance under changes of coordinates (thus geometric nature of $ \epsilon $).
\item The definitions are algebraic and thus work for any \emph{algebraic field} (or even ring) in place of $ \mathbb{C} $. {\color{BrickRed}It is true}.
\item What if $ \epsilon (A_1, A_2, \cdots , A_N) = 0 $? What does this mean geometrically? {\color{BrickRed}Suggestion: topic for later (would imply many geometric properties, so maybe another paper?)}.

\end{itemize}
\fi

\section{Alternative expression and cases $N=4,5$}
In this appendix, we present an alternative general form of the polydeterminant $\epsilon$-function in terms of traces and explicit rather involved specific expressions for $N=4,5$. Namely, these two cases could be relevant when discussing anomalous breaking terms involving the charm and bottom mesons.

The polydeterminant can be written down as:%
\begin{equation}
\epsilon(A_{1},A_{2},...,A_{N})=\sum_{\substack{n_{1},..,n_{N}\geq
0\\n_{1}+2n_{2}+...+Nn_{N}=N\text{ }}}\frac{(-1)^{n_{1}+n_{2}+...n_{N}+N}}%
{N!}Y^{n_{1}n_{2}...n_{N}}%
\end{equation}
with%
\begin{align}
Y^{n_{1}n_{2}\cdots n_{N}} &  =\mathrm{Tr}\left(  A_{1}\right)  \mathrm{Tr}%
\left(  A_{2}\right)  \cdot...\cdot\mathrm{Tr}\left(  A_{n_{1}}\right)
\nonumber\\
&  \mathrm{Tr}\left(  A_{n_{1}+1}A_{n_{1}+2}\right)  \mathrm{Tr}\left(
A_{n_{1}+3}A_{n_{1}+4}\right)  \cdot...\cdot\mathrm{Tr}\left(  A_{n_{1}%
+2n_{2}-1)}A_{n_{1}+2n_{2}}\right)  \nonumber\\
&  \mathrm{Tr}\left(  A_{n_{1}+2n_{2}+1}A_{n_{1}+2n_{2}+2}A_{n_{1}+2n_{2}%
+1}\right)  \cdot....\text{ distinct terms,}%
\end{align}
where `distinct terms' refer to those permutations that deliver, for generic arbitrary matrices, different results. 
For instance,
\begin{equation}
Y^{N0....0}=\mathrm{Tr}\left(  A_{1}\right)  \mathrm{Tr}\left(  A_{2}\right)
\cdot...\cdot\mathrm{Tr}\left(  A_{N}\right)
\end{equation}
contains one single term (all permutations lead to the very same term). On the
other hand:%
\begin{align}
Y^{(N-2)1....0}  & =\mathrm{Tr}\left(  A_{1}\right)  \mathrm{Tr}\left(
A_{2}\right)  \cdot...\cdot\mathrm{Tr}\left(  A_{N-2}\right)  \mathrm{Tr}%
\left(  A_{N-1}A_{N}\right)  \nonumber\\
& +\mathrm{Tr}\left(  A_{N}\right)  \mathrm{Tr}\left(  A_{2}\right)
\cdot...\cdot\mathrm{Tr}\left(  A_{N-2}\right)  \mathrm{Tr}\left(
A_{N-1}A_{1}\right)  +...
\end{align}
contains $N!/(N-2)!=N(N-1)$ distinct terms. 
The last term is:%
\begin{equation}
Y^{00...1}=\mathrm{Tr}\left(  A_{1}A_{2}...A_{N}\right)  +\mathrm{Tr}\left(
A_{2}A_{1}...A_{N}\right)  +...
\end{equation}
contains $N!/N=(N-1)!$ distinct terms.

In general, the number of distinct terms within $Y^{n_{1}n_{2}\cdots n_{N}}$
amounts to%
\begin{equation}
N!\left\vert C^{n_{1}n_{2}\cdots n_{N}}\right\vert \text{ ,}%
\end{equation}
where the coefficients \(C^{n_{1}n_{2}\cdots n_{N}}\) are given in \eqref{coeff}.

Next, for $N=4$, the explicit expression reads
   \begin{align*}
   \epsilon(A,B,C,D)= \frac{1}{24}\Big(\text{Tr}(A)\text{Tr}(B)\text{Tr}(C)\text{Tr}(D)-
\Big(\text{Tr}(A)\text{Tr}(B)\text{Tr}(CD)+\\\text{Tr}(A)\text{Tr}(C)\text{Tr}(BD)+\text{Tr}(A)\text{Tr}(D)\text{Tr}(BC)+\\
\text{Tr}(B)\text{Tr}(C)\text{Tr}(AD)+\text{Tr}(B)\text{Tr}(D)\text{Tr}(AC)+\text{Tr}(C)\text{Tr}(D)\text{Tr}(AB)\Big)+\\
   \text{Tr}(AB)\text{Tr}(CD)+ \text{Tr}(AD)\text{Tr}(BC)+ \text{Tr}(AC)\text{Tr}(BD)+\\
   \text{Tr}(A)\text{Tr}(BCD)+\text{Tr}(A)\text{Tr}(BDC)+\text{Tr}(B)\text{Tr}(CDA)+\text{Tr}(B)\text{Tr}(CAD)+\\
  \text{Tr}(C)\text{Tr}(DAB)+ \text{Tr}(C)\text{Tr}(DBA)+ \text{Tr}(D)\text{Tr}(ABC)+\text{Tr}(D)\text{Tr}(ACB) -\\
 \Big(  \text{Tr}(ABCD)+\text{Tr}(ABDC)+\text{Tr}(ACBD)+\text{Tr}(ACDB)+  \text{Tr}(ADBC)+  \text{Tr}(ADCB)\Big)\Big)\,.   
    \end{align*}
It reduces to the following known relation in the limit of $A=B=C=D$
   \begin{align*}
    \det{A}=\frac{1}{24}\Big(\text{Tr}(A)^4-6\text{Tr}(A^2)(\text{Tr}(A))^2+3(\text{Tr}(A^2))^2+8\text{Tr}(A)\text{Tr}(A^3)-6\text{Tr}(A^4)\Big)\,.
    \end{align*}   

 For the case of $N=5$, the $\epsilon$ function acting on the $5\times 5$ complex matrices reads

  \begin{align*}
  \epsilon(A,B,C,D,E)=\frac{1}{120} \Big[ \text{Tr}(ABCDE)
    -     \text{Tr}(AB) \text{Tr}(C)\text{Tr}(D)\text{Tr}(E)
    - 
    \text{Tr}(AC) \, \text{Tr}(B)\text{Tr}(D)\text{Tr}(E)
    - \\
     \text{Tr}(AD) \, \text{Tr}(B)\text{Tr}(C)\text{Tr}(E)
    - \, \text{Tr}(AE) \, \text{Tr}(B)\text{Tr}(C)\text{Tr}(D)
    -
     \, \text{Tr}(BD) \, \text{Tr}(C)\text{Tr}(A)\text{Tr}(E)
    -  \, \text{Tr}(BE) \, \text{Tr}(C)\text{Tr}(A)\text{Tr}(D)
    - \\
    \, \text{Tr}(BC) \, \text{Tr}(D)\text{Tr}(A)\text{Tr}(E)
    - \, \text{Tr}(CD) \, \text{Tr}(B)\text{Tr}(A)\text{Tr}(E)
    -  \, \text{Tr}(CE) \, \text{Tr}(A)\text{Tr}(B)\text{Tr}(D)
    -  \, \text{Tr}(DE) \, \text{Tr}(A)\text{Tr}(B)\text{Tr}(C)
    +  \\
    \, \text{Tr}(ABC) \, \text{Tr}(D) \, \text{Tr}(E)+  \, \text{Tr}(ACB) \, \text{Tr}(D) \, \text{Tr}(E)+  \, \text{Tr}(ABD) \, \text{Tr}(C) \, \text{Tr}(E)+ \, \text{Tr}(ADB) \, \text{Tr}(C) \, \text{Tr}(E)+\\
    \, \text{Tr}(ABE) \, \text{Tr}(C) \, \text{Tr}(D)+ \, \text{Tr}(AEB) \, \text{Tr}(C) \, \text{Tr}(D)+ \, \text{Tr}(ACD) \, \text{Tr}(B) \, \text{Tr}(E)+ \, \text{Tr}(ADC) \, \text{Tr}(B) \, \text{Tr}(E)+
    \\
     \, \text{Tr}(ACE) \, \text{Tr}(B) \, \text{Tr}(D)+ \, \text{Tr}(AEC) \, \text{Tr}(B) \, \text{Tr}(D)+ \, \text{Tr}(ADE) \, \text{Tr}(B) \, \text{Tr}(C)+\, \text{Tr}(AED) \, \text{Tr}(B) \, \text{Tr}(C)+\\
      \, \text{Tr}(BCD) \, \text{Tr}(A) \, \text{Tr}(E)+  \, \text{Tr}(BDC) \, \text{Tr}(A) \, \text{Tr}(E)+  \, \text{Tr}(BCE) \, \text{Tr}(A) \, \text{Tr}(D)+\text{Tr}(BEC) \, \text{Tr}(A) \, \text{Tr}(D)+\\
      \, \text{Tr}(BDE) \, \text{Tr}(A) \, \text{Tr}(C)+\, \text{Tr}(BED) \, \text{Tr}(A) \, \text{Tr}(C)+    \, \text{Tr}(CDE) \, \text{Tr}(A) \, \text{Tr}(B)+ \, \text{Tr}(CED) \, \text{Tr}(A) \, \text{Tr}(B)+\\
     \text{Tr}(A) \, (\text{Tr}(BC)) (\text{Tr}(DE)) \,  + \text{Tr}(A) \, (\text{Tr}(BD)) (\text{Tr}(CE)) \,  + \text{Tr}(A) \, (\text{Tr}(BE)) (\text{Tr}(CD)) \, 
    +\\
    \text{Tr}(B) \, \text{Tr}(AC)\text{Tr}(DE) \,  + \text{Tr}(B) \, \text{Tr}(AD) \text{Tr}(CE) \,  + \text{Tr}(B) \, \text{Tr}(AE) \text{Tr}(CD) \,+ \\
    \text{Tr}(C) \, \text{Tr}(AB) \text{Tr}(DE) \,  + \text{Tr}(C) \, \text{Tr}(AD)\text{Tr}(BE) \,  + \text{Tr}(C) \, \text{Tr}(AE) \text{Tr}(BD) \,+ \\
    \text{Tr}(D) \, \text{Tr}(AB) \text{Tr}(CE) \,  + \text{Tr}(D) \, \text{Tr}(AC) \text{Tr}(BE) \,  + \text{Tr}(D) \, \text{Tr}(AE) \text{Tr}(BC) \,+ \\
    \text{Tr}(E) \, \text{Tr}(AB) \text{Tr}(CD) \,  + \text{Tr}(E) \, \text{Tr}(AC) \text{Tr}(BD) \,  + \text{Tr}(E) \, \text{Tr}(AD) \text{Tr}(BC) \,-\\
    \text{Tr}(A) \, \Big(  \text{Tr}(EBCD)+\text{Tr}(EBDC)+\text{Tr}(ECBD)+\text{Tr}(ECDB)+  \text{Tr}(EDBC)+  \text{Tr}(EDCB)\Big) -\\
     \text{Tr}(B) \, \Big(  \text{Tr}(AECD)+\text{Tr}(AEDC)+\text{Tr}(ACED)+\text{Tr}(ACDE)+  \text{Tr}(ADEC)+  \text{Tr}(ADCE)\Big) -\\
     \text{Tr}(C) \, \Big(  \text{Tr}(ABED)+\text{Tr}(ABDE)+\text{Tr}(AEBD)+\text{Tr}(AEDB)+  \text{Tr}(ADBE)+  \text{Tr}(ADEB)\Big) -\\              \text{Tr}(D) \, \Big(  \text{Tr}(ABCE)+\text{Tr}(ABEC)+\text{Tr}(ACBE)+\text{Tr}(ACEB)+  \text{Tr}(AEBC)+  \text{Tr}(AECB)\Big) -\\
    \text{Tr}(E) \, \Big(  \text{Tr}(ABCD)+\text{Tr}(ABDC)+\text{Tr}(ACBD)+\text{Tr}(ACDB)+  \text{Tr}(ADBC)+  \text{Tr}(ADCB)\Big) 
    - \\
        \Big( \, \text{Tr}(ABC) \, \text{Tr}(D E)+  \, \text{Tr}(ACB) \, \text{Tr}(D E)+  \, \text{Tr}(ABD) \, \text{Tr}(C E)+ \, \text{Tr}(ADB) \, \text{Tr}(C E)\Big)-\\
     \Big(  \, \text{Tr}(ABE) \, \text{Tr}(C D)+ \, \text{Tr}(AEB) \, \text{Tr}(C D)+ \, \text{Tr}(ACD) \, \text{Tr}(B E)+ \, \text{Tr}(ADC) \, \text{Tr}(B E)\Big)-
    \\
     \Big( \, \text{Tr}(ACE) \, \text{Tr}(B D)+ \, \text{Tr}(AEC) \, \text{Tr}(B D)+ \, \text{Tr}(ADE) \, \text{Tr}(B C)+\, \text{Tr}(AED) \, \text{Tr}(B C)\Big)-\\
     \Big( \, \text{Tr}(BCD) \, \text{Tr}(A E)+  \, \text{Tr}(BDC) \, \text{Tr}(A E)+  \, \text{Tr}(BCE) \, \text{Tr}(A D)+\text{Tr}(BEC) \, \text{Tr}(A D)\Big)-\\
 \Big(\, \text{Tr}(BDE) \, \text{Tr}(A C)+\, \text{Tr}(BED) \, \text{Tr}(A C)+    \, \text{Tr}(CDE) \, \text{Tr}(A B)+ \, \text{Tr}(CED) \, \text{Tr}(A B)\Big)+\\
    \text{Tr}(A B C D E) + \text{Tr}(A B C E D) +\text{Tr}(A B D C E) +  \text{Tr}(A B D E C) + \text{Tr}(A B E C D) + \text{Tr}(A B E D C)+ \\
\text{Tr}(A C B D E) +\text{Tr}(A C B E D) + \text{Tr}(A C D B E) +\text{Tr}(A C D E B) +\text{Tr}(A C E B D) +\text{Tr}(A C E D B)+ \\
\text{Tr}(A D B C E) +\text{Tr}(A D B E C) + \text{Tr}(A D C B E) +\text{Tr}(A D C E B) + \text{Tr}(A D E B C) + \text{Tr}(A D E C B)+ \\
\text{Tr}(A E B C D) + \text{Tr}(A E B D C) +\text{Tr}(A E C B D) + \text{Tr}(A E C D B) + \text{Tr}(A E D B C) + \text{Tr}(A E D C B)    \Big)
\end{align*}
which reduces to the following relation for the $5\times 5$ matrices in the limit of $A=B=C=D=E$
\begin{align}\nonumber
    \det{A} &= \frac{1}{120} \Big( 
    \text{Tr}(A)^5 
    - 10 \, \text{Tr}(A^2) \, \text{Tr}(A)^3 
    + 20 \, \text{Tr}(A^3) \, \text{Tr}(A)^2 + 15 \, (\text{Tr}(A^2))^2 \, \text{Tr}(A) 
    -
    \\
   &  \qquad\qquad \qquad 30 \, \text{Tr}(A^4) \, \text{Tr}(A) 
    - 20 \, \text{Tr}(A^2) \, \text{Tr}(A^3) 
    + 24 \, \text{Tr}(A^5) 
    \Big)
\end{align}

\newpage

\bibliographystyle{alpha}
\bibliography{main}

\newcommand{\etalchar}[1]{$^{#1}$}
\begin{thebibliography}{PKW{\etalchar{+}}13}

\bibitem[Ale38]{Alexandroff1938}
A.~Alexandroff.
\newblock Zur theorie der gemischten volumina von konvexen körpern. iv.
\newblock {\em Matematicheskii Sbornik}, 3(45):227--251, 1938.
\newblock German summary.

\bibitem[AS23]{ayyer2023combinatorial}
Arvind Ayyer and Naren Sundaravaradan.
\newblock Combinatorial proofs of multivariate cayley--hamilton theorems.
\newblock {\em Linear Algebra and its Applications}, 661:247--269, 2023.

\bibitem[Bap89]{BAPAT1989107}
R.B. Bapat.
\newblock Mixed discriminants of positive semidefinite matrices.
\newblock {\em Linear Algebra and its Applications}, 126:107--124, 1989.

\bibitem[Bap15]{Bapat2015CayleyHamiltonTF}
Ravindra~B. Bapat.
\newblock Cayley-hamilton theorem for mixed discriminants.
\newblock 2015.

\bibitem[BF84]{Backhouse:1984vn}
N.~B. Backhouse and A.~G. Fellouris.
\newblock {ON THE SUPERDETERMINANT FUNCTION FOR SUPERMATRICES}.
\newblock {\em J. Phys. A}, 17:1389--1395, 1984.

\bibitem[BPST75]{Belavin:1975fg}
A.~A. Belavin, Alexander~M. Polyakov, A.~S. Schwartz, and Yu.~S. Tyupkin.
\newblock {Pseudoparticle Solutions of the Yang-Mills Equations}.
\newblock {\em Phys. Lett. B}, 59:85--87, 1975.

\bibitem[CCD{\etalchar{+}}13]{cattani2013mixed}
Eduardo Cattani, Mar{\'\i}a~Ang{\'e}lica Cueto, Alicia Dickenstein, Sandra Di~Rocco, and Bernd Sturmfels.
\newblock Mixed discriminants.
\newblock {\em Mathematische Zeitschrift}, 274(3):761--778, 2013.

\bibitem[Dre05]{drensky2005computing}
Vesselin Drensky.
\newblock Computing with matrix invariants.
\newblock {\em arXiv preprint math/0506614}, 2005.

\bibitem[EGR15]{Eshraim:2014eka}
Walaa~I. Eshraim, Francesco Giacosa, and Dirk~H. Rischke.
\newblock {Phenomenology of charmed mesons in the extended Linear Sigma Model}.
\newblock {\em Eur. Phys. J. A}, 51(9):112, 2015.

\bibitem[FJS05]{Fariborz:2005gm}
Amir~H. Fariborz, Renata Jora, and Joseph Schechter.
\newblock {Toy model for two chiral nonets}.
\newblock {\em Phys. Rev. D}, 72:034001, 2005.

\bibitem[FM09]{Forrester2009}
Peter~J. Forrester and Anthony Mays.
\newblock Pfaffian point process for the gaussian real generalised eigenvalue problem.
\newblock {\em Linear Algebra and its Applications}, 430(3):868--892, 2009.

\bibitem[FMS16]{florentin2016characterization}
D~Florentin, V~Milman, and R~Schneider.
\newblock A characterization of the mixed discriminant.
\newblock {\em Proceedings of the American Mathematical Society}, 144(5):2197--2204, 2016.

\bibitem[FS17]{Fukushima:2017csk}
Kenji Fukushima and Vladimir Skokov.
\newblock {Polyakov loop modeling for hot QCD}.
\newblock {\em Prog. Part. Nucl. Phys.}, 96:154--199, 2017.

\bibitem[FW15]{frenkel2015classical}
P{\'e}ter~E Frenkel and Mih{\'a}ly Weiner.
\newblock Classical information storage in an n-level quantum system.
\newblock {\em Communications in Mathematical Physics}, 340(2):563--574, 2015.

\bibitem[Gel89]{gelfand1989lectures}
I.M. Gelfand.
\newblock {\em Lectures on Linear Algebra}.
\newblock Dover Books on Mathematics Series. Dover Publications, 1989.

\bibitem[Gia18]{Giacosa:2017ojs}
Francesco Giacosa.
\newblock {Revisiting the axial anomaly for light mesons and baryons}.
\newblock {\em PoS}, Hadron2017:045, 2018.

\bibitem[GJP24]{Giacosa:2023fdz}
Francesco Giacosa, Shahriyar Jafarzade, and Robert~D. Pisarski.
\newblock {Anomalous interactions between mesons with nonzero spin and glueballs}.
\newblock {\em Phys. Rev. D}, 109(7):L071502, 2024.

\bibitem[GKJ24]{Giacosa:2024epf}
Francesco Giacosa, P\'eter Kov\'acs, and Shahriyar Jafarzade.
\newblock {Ordinary and exotic mesons in the extended Linear Sigma Model}.
\newblock 7 2024.

\bibitem[GKK{\etalchar{+}}25]{Giacosa:2024orp}
Francesco Giacosa, Gy\H{o}z\H{o} Kov\'acs, P\'eter Kov\'acs, Robert~D. Pisarski, and Fabian Rennecke.
\newblock {Anomalous U(1)A couplings and the Columbia plot}.
\newblock {\em Phys. Rev. D}, 111(1):016014, 2025.

\bibitem[GKP18]{Giacosa:2017pos}
Francesco Giacosa, Adrian Koenigstein, and Robert~D. Pisarski.
\newblock {How the axial anomaly controls flavor mixing among mesons}.
\newblock {\em Phys. Rev. D}, 97(9):091901, 2018.

\bibitem[GKZ92]{GELFAND1992226}
I.M Gelfand, M.M Kapranov, and A.V Zelevinsky.
\newblock Hyperdeterminants.
\newblock {\em Advances in Mathematics}, 96(2):226--263, 1992.

\bibitem[GML60]{Gell-Mann:1960mvl}
Murray Gell-Mann and M~Levy.
\newblock {The axial vector current in beta decay}.
\newblock {\em Nuovo Cim.}, 16:705, 1960.

\bibitem[Gur04]{GURVITS2004448}
Leonid Gurvits.
\newblock Classical complexity and quantum entanglement.
\newblock {\em Journal of Computer and System Sciences}, 69(3):448--484, 2004.
\newblock Special Issue on STOC 2003.

\bibitem[Hei32]{Heisenberg:1932dw}
W.~Heisenberg.
\newblock {On the structure of atomic nuclei}.
\newblock {\em Z. Phys.}, 77:1--11, 1932.

\bibitem[LM15]{Lehman:2015via}
Landon Lehman and Adam Martin.
\newblock {Hilbert Series for Constructing Lagrangians: expanding the phenomenologist's toolbox}.
\newblock {\em Phys. Rev. D}, 91:105014, 2015.

\bibitem[Mos89]{Mosel:1989jf}
U.~Mosel.
\newblock {\em {Fields, symmetries, and quarks}}.
\newblock Springer-Verlag, 1989.

\bibitem[N{\etalchar{+}}24]{ParticleDataGroup:2024cfk}
S.~Navas et~al.
\newblock {Review of particle physics}.
\newblock {\em Phys. Rev. D}, 110(3):030001, 2024.

\bibitem[NGE98]{Neufeld:1998js}
H.~Neufeld, J.~Gasser, and G.~Ecker.
\newblock {The one loop functional as a Berezinian}.
\newblock {\em Phys. Lett. B}, 438:106--114, 1998.

\bibitem[Ott13]{Ottaviani2013}
Giorgio Ottaviani.
\newblock Introduction to the hyperdeterminant and to the rank of multidimensional matrices.
\newblock 2013.

\bibitem[Pan87]{AAPanov1987}
A~A Panov.
\newblock On some properties of mixed discriminants.
\newblock {\em Mathematics of the USSR-Sbornik}, 56(2):279, feb 1987.

\bibitem[PG17]{Parganlija:2016yxq}
Denis Parganlija and Francesco Giacosa.
\newblock {Excited Scalar and Pseudoscalar Mesons in the Extended Linear Sigma Model}.
\newblock {\em Eur. Phys. J. C}, 77(7):450, 2017.

\bibitem[PKW{\etalchar{+}}13]{Parganlija:2012fy}
Denis Parganlija, Peter Kovacs, Gyorgy Wolf, Francesco Giacosa, and Dirk~H. Rischke.
\newblock {Meson vacuum phenomenology in a three-flavor linear sigma model with (axial-)vector mesons}.
\newblock {\em Phys. Rev. D}, 87(1):014011, 2013.

\bibitem[PR20]{Pisarski:2019upw}
Robert~D. Pisarski and Fabian Rennecke.
\newblock {Multi-instanton contributions to anomalous quark interactions}.
\newblock {\em Phys. Rev. D}, 101(11):114019, 2020.

\bibitem[PR24]{Pisarski:2024esv}
Robert~D. Pisarski and Fabian Rennecke.
\newblock {Conjectures about the Chiral Phase Transition in QCD from Anomalous Multi-Instanton Interactions}.
\newblock {\em Phys. Rev. Lett.}, 132(25):251903, 2024.

\bibitem[Pro76]{PROCESI1976306}
C~Procesi.
\newblock The invariant theory of n × n matrices.
\newblock {\em Advances in Mathematics}, 19(3):306--381, 1976.

\bibitem[Pro20]{procesi2020tensor}
Claudio Procesi.
\newblock Tensor fundamental theorems of invariant theory.
\newblock {\em arXiv preprint arXiv:2011.10820}, 2020.

\bibitem[RSS81]{Rosenzweig:1981cu}
C.~Rosenzweig, A.~Salomone, and J.~Schechter.
\newblock {A Pseudoscalar Glueball, the Axial Anomaly and the Mixing Problem for Pseudoscalar Mesons}.
\newblock {\em Phys. Rev. D}, 24:2545--2548, 1981.

\bibitem[Str09]{strang2009introduction}
Gilbert Strang.
\newblock {\em Introduction to Linear Algebra}.
\newblock Wellesley-Cambridge Press, 4th edition, 2009.

\bibitem[tH76a]{tHooft1976}
G.~'t~Hooft.
\newblock Computation of the quantum effects due to a four-dimensional pseudoparticle.
\newblock {\em Phys. Rev. D}, 14:3432--3450, Dec 1976.

\bibitem[tH76b]{tHooft:1976snw}
Gerard 't~Hooft.
\newblock {Computation of the Quantum Effects Due to a Four-Dimensional Pseudoparticle}.
\newblock {\em Phys. Rev. D}, 14:3432--3450, 1976.
\newblock [Erratum: Phys.Rev.D 18, 2199 (1978)].

\bibitem[tH76c]{tHooft:1976rip}
Gerardus 't~Hooft.
\newblock {Symmetry breaking through Bell-Jackiw anomalies}.
\newblock {\em Phys. Rev. Lett.}, 37:8--11, 7 1976.

\bibitem[tH86]{tHooft:1986ooh}
Gerardus 't~Hooft.
\newblock {How instantons solve the $U(1)$ problem}.
\newblock {\em Phys. Rept.}, 142(6):357--387, 9 1986.

\bibitem[tH99]{tHooft:1999cta}
Gerard 't~Hooft.
\newblock {The Physics of instantons in the pseudoscalar and vector meson mixing}.
\newblock 3 1999.

\bibitem[Wig37]{Wigner:1936dx}
E.~Wigner.
\newblock {On the Consequences of the Symmetry of the Nuclear Hamiltonian on the Spectroscopy of Nuclei}.
\newblock {\em Phys. Rev.}, 51:106--119, 1937.

\bibitem[WN11]{WimmerPfaffian}
E.~Wimmer and M.~A. Nielsen.
\newblock Efficient numerical computation of the pfaffian for dense skew-symmetric matrices.
\newblock {\em arXiv}, 2011.

\bibitem[Zee16]{Zee:2016fuk}
Anthony Zee.
\newblock {\em {Group Theory in a Nutshell for Physicists}}.
\newblock Princeton University Press, USA, 3 2016.

\end{thebibliography}

\end{document}